\documentstyle[preprint,prl,aps]{revtex}
\begin{document}

\draft

\title{C-axis Electronic Raman Scattering in 
Bi$_2$Sr$_2$CaCu$_2$O$_{8+\delta}$}

\author{H.L. Liu,$^1$ G. Blumberg,$^1$ M.V. Klein,$^1$
P. Guptasarma,$^2$ and D.G. Hinks$^2$}

\address{$^1$ Department of Physics and Science and Technology Center for 
Superconductivity,\\ University of Illinois at Urbana-Champaign, Urbana, 
Illinois 61801-3080\\
$^2$ Materials Science Division and Science and Technology Center for 
Superconductivity,\\ Argonne National Laboratory, Argonne, Illinois 60439}

\date{\today}
\maketitle

\begin{abstract}
We report a $c$-axis-polarized electronic Raman scattering study of 
Bi$_2$Sr$_2$CaCu$_2$O$_{8+\delta}$ single crystals. 
In the normal state, a resonant electronic continuum extends to 1.5 eV and
gains significant intensity as the incoming photon energy increases. 
In the superconducting state, a coherence 2$\Delta$ peak appears around 
50 meV, with a suppression
of the scattering intensity at frequencies below the peak position. 
The peak energy, which is higher than that 
seen with in-plane polarizations, signifies distinctly different dynamics 
of quasiparticle excitations created with out-of-plane polarization. 
\end{abstract}


\narrowtext

One of the peculiar aspects of the high-$T_c$ cuprates is the incoherent
nature of the charge transport perpendicular to the CuO$_2$ 
planes\cite{Cooper94}. 
Early on it was established experimentally that the normal-state
in-plane resistivity typically varies linearly with temperature, 
whereas the out-of-plane resistivity almost universally displays 
semiconducting behavior\cite{Ito91}. 
The $c$-axis optical conductivity of the most 
cuprates\cite{Tajima93,Timusk95} shows
an electronic background with a very large scattering rate -- that is
the mean free path appears to be 
less than the lattice spacing. 
These results suggest that there is no coherent electronic 
transport in the $c$-direction: all motions are inelastic.
A number of models have been proposed to explain the 
important mechanisms contributing to $c$-axis transport in high-$T_c$ 
superconductors, including localization along the $c$-axis\cite{Kotliar91},
carrier confinement in the resonating valence bond (RVB) 
theory\cite{Wheatley88}, and interlayer tunneling (ILT) 
or hopping\cite{Anderson91,Leggett96}. However, there is currently no
consensus as to the clear picture of the origin of incoherent $c$-axis 
transport.

Among all cuprate superconductors, the double-layered 
Bi$_2$Sr$_2$CaCu$_2$O$_{8+\delta}$ (Bi-2212) crystal 
is the most decoupled material.
In Bi-2212, the ratio of the out-of-plane to in-plane
resistivity $\rho_c/\rho_{ab}$ can be as high as 10$^5$\cite{Martin88}. 
Optical response shows that the $c$-axis reflectivity of Bi-2212 is highly 
insulating, while the in-plane reflectivity is metallic\cite{Tajima93}.   
The corresponding plasma frequency anisotropy in Bi-2212, 
$\omega_{p ab} / \omega_{p c}$ 
$>$ 100, is substantially larger than that observed in other cuprates. 
These data provide evidence that the $c$-axis transport of Bi-2212 is 
incoherent, with the extremely small energy scale set by the
hopping interaction between the adjacent CuO$_2$ bilayers.
In the superconducting state, the two-dimensional character of Bi-2212 
also manifests itself strongly in the penetration depth measurements.
The $c$-axis penetration depth is extremely large ($\lambda_c \approx$ 100 
$\mu$m)\cite{Cooper90}.
The large anisotropy between $\lambda_{ab}$ and $\lambda_c$
of Bi-2212 was shown to be best described within a picture of
strongly superconducting CuO$_2$ layers weakly coupled by Josephson 
interaction along the $c$-axis\cite{Jacobs95}.

The purpose of this study is to investigate, in the context of electronic
Raman scattering spectroscopy, the role of $c$-axis polarizability in 
the Bi-2212 cuprates. Raman scattering has been proved to 
be a valuable technique for understanding the quasiparticle dynamics  on 
different regions of the Fermi surface in the cuprate systems 
by orienting incoming and outgoing photon 
polarizations\cite{Devereaux97}. 
The electronic Raman spectra polarized in the $ab$-plane of Bi-2212 have been extensively 
studied\cite{Kendziora95}. In contrast,
Raman data on the electronic scattering of Bi-2212 for photons polarized
along the $c$-axis are 
rare\cite{Boekholt91}. 
Our new results show that the electronic continuum
in $zz$-polarization does exist and is not small. Notably, this  
continuum intensity resonates towards near ultraviolet (UV) 
photon excitation. 
Below $T_c$, there is a measurable superconductivity-induced redistribution 
of the $zz$-polarized continuum and the presence of 
a 2$\Delta$ peak-like feature, similar to those observed in the 
$ab$-plane Raman response.

Single crystals of Bi-2212 were grown near-stoichiometric using a
solvent-free Floating Zone process in a double-mirror image furnace modified 
for very slow growth. In this letter, we used an as-grown, un-annealed single
crystal of dimensions $5 \times 1 \times 0.5$ mm$^3$ with a 
superconducting transition onset at 87 K (dc magnetization) and 
onset-to-saturation midpoint at 85 K. The Raman 
measurements were performed on two faces of the crystal. One face 
(labeled I in Fig.~1) contains both 
the $c$-axis and either the $a$- or $b$- direction. 
The second face, face II, 
provided the $a$- and $b$-axis response. 
We have also studied another sample ($T_c$ = 93 K, 
$\Delta{T_c}$ = 1.5 K) with less surface quality of face I and obtained similar results.
Throughout this study, $x$ and $y$ are indexed along the Bi-O 
bonds, rotated by $45^{\circ}$ with respect to the Cu-O bonds. All symmetries 
refer to a tetragonal $D_{4h}$ point group.

The low-frequency Raman spectra were taken in
pseudobackscattering geometry with $\hbar\omega_i$ = 1.92 eV
photons from a Kr$^+$ laser. The laser excitation of less 
than 10 W/cm$^2$ was focused into a 50 
$\mu$m diameter spot on the sample surface. The 
temperatures referred to in this paper are the nominal temperatures inside 
the cryostat. The spectra were analyzed by a triple grating spectrometer with
a liquid-nitrogen cooled charge-coupled device detector.    
To investigate further the resonance property, 
we have used several excitation lines from Ar$^+$ and Kr$^+$ 
lasers ranging from visible red to near UV. All the Raman spectra were 
corrected for the spectral response of the spectrometer and detector, 
the optical absorption of the sample as well as the refraction at the 
sample-gas interface\cite{Reznik92}.

The imaginary parts of the $c$-axis-polarized Raman response functions, 
obtained by dividing
the original spectra by the Bose-Einstein thermal factor, are shown in  
Fig.~1 for two different temperatures, 100 K ($T > {T_c}$) and 5 K
($T \ll {T_c}$). In the normal state, the most prominent features of 
the spectra are the electronic continuum and 
several $q \approx$ 0 Raman allowed phonon modes, whose overall character
is in good agreement with that reported 
previously\cite{Denisov89,Liu92,Kakihana96}.
We focus on the temperature behavior of the 
electronic Raman scattering response. Well below $T_c$, 
it can seen in Fig.~1 that for the $zz$ continuum there is a loss of 
scattering strength at low frequencies which redistributes into the weak and 
broad peak at higher frequencies. 
Above 700 cm$^{-1}$, the 5 K and 100 K spectra appear to be essentially 
identical. These data were reproducibly observed at 
three different spots on the sample.
We emphasize that the redistribution of the scattering 
intensity itself is not of phononic origin. The low energy (red) excitation is used primarily to 
reduce the intensity of phononic scattering. Furthermore, a similar feature
has been reported for other less anisotropic members of 
the high-$T_c$ superconductors, including
YBa$_2$Cu$_3$O$_{7-\delta}$ (YBCO)\cite{McCarty90} and 
NdBa$_2$Cu$_3$O$_{7-\delta}$\cite{Misochko97}.

In order to observe the superconductivity-induced 
redistribution of the electronic continuum better, in the inset of Fig.~1
we present the change between the normal and superconducting spectra in an 
enhanced manner, where the 5 K spectrum is normalized by (1) dividing by 
the 100 K spectrum (top curve), (2) subtracting the 100 K spectrum (middle 
curve), and (3) the difference of Raman spectra, as in (2),
after first subtracting the phonon contributions (bottom curve). 
We believe that sharp features in difference spectra are due to 
temperature dependence of phononic scatterings.
It is nevertheless clear that in all cases below $T_c$ 
a broad peak forms in the electronic continuum which
is accompanied by reduced scattering at the lowest Raman shift.
The difference vanishes at sufficiently high 
frequencies.

It is instructive to compare the $c$-axis electronic continuum with results 
for other scattering configurations, which are shown in Fig.~2. 
As can be seen in the Fig.~2(a), a depolarized $zx$ spectrum 
has an even smaller continuum intensity compared with that from 
the $zz$ component. 
Furthermore, the normal and superconducting spectra are indistinguishable. 
In contrast, the superconducting transition 
leads to the redistribution of the continuum into a broad peak 
in $xx$ polarization (Fig.~2(b)). It is interesting to note that
the out-of-plane and in-plane $xx$ spectra (Fig.~2(c)) look almost identical. 
We have also found that there is
an $x$-$y$ anisotropy clearly demonstrated by the phonon modes between 
in-plane $xx$ and $yy$ (Fig.~2(f)) polarizations\cite{Liu92}. 
Referring to the Fig.~2(e),
the $B_{1g}$ contribution is predominant in $xy$ geometry, and gives an electronic 
continuum that is much stronger than that in any other polarization.
Below $T_c$, the strong suppression of the continuum is observed, and the 
low-frequency intensity varies roughly as $\omega^3$, while it is quite 
linear in $x^{\prime}y^{\prime}$ (Fig.~2(d)) polarization 
($B_{2g} + A_{2g}$ symmetry). At the same time, the magnitude of 
superconductivity-induced peak is much less intense in $B_{2g} + A_{2g}$ 
symmetry than that found in $B_{1g}$ symmetry.
Such $\omega$ dependences in both scattering geometries are consistent with
an order parameter of $d$-wave symmetry [$d(x^2 - y^2)$ when referred to Cu-O
bonds]\cite{Devereaux94}.

The results presented in Figs.~1 and~2 clearly show that
the intensity of the $c$-polarized continuum is not 
negligible compared with that of the in-plane symmetries. 
We regard these observations as truly extraordinary, for $c$-axis transport 
is incoherent. To examine what microscopic origins 
might produce the $zz$ continuum in Bi-2212, we first discuss our data
in terms of the conventional model of light scattering from a 
superconductor\cite{Devereaux94,Klein84}. 
In this model, the strength of the electronic Raman 
scattering is proportional to the square of the Raman vertex. For 
nonresonant excitation, the Raman vertex at a point $k$ in reciprocal space
is given approximately by the 
inverse effective mass (the curvature of the energy band dispersion), 
$\gamma_{\rm i j} \propto {1 \over {m^*_{\rm i j}}} 
\propto {{\partial^2 \epsilon(k)} \over {\partial k_i \partial k_j}}$. 
In Bi-2212, both transport\cite{Tajima93,Martin88} and band structure 
calculations\cite{Pickett89} reflect the fact that
the $c$-axis effective 
mass is in general very heavy. Consequently, 
the $c$-axis nonresonant Raman scattering should be truly small.
We see, instead, an ordinary size for the electronic 
Raman scattering intensity along the
$c$-direction, and suggest that this is a result of a resonance Raman 
vertex.

To illustrate this consideration, we show in Fig.~3 on an expanded 
scale room-temperature $c$-axis Raman spectra 
excited with photon energies between 1.92 and 3.05 eV. For the latter 
incident energy, we followed the continuum to 1.5 eV Raman shift. 
The continuum intensity exhibits a dramatic increase 
when the excitation approaches the UV region.
This is more clearly seen in the inset of Fig.~3 where we
plot the resonance profile for the $zz$- and $xx$-polarized continua.  
The continuum intensity was measured at around 2000 cm$^{-1}$ and 
normalized to its value at 3.05 eV. We note that the $zz$ continuum gains 
intensity by a factor of $\approx$ 5 towards UV excitation,
while the change of the $xx$ continuum intensity is rather modest. 
Due to the limited number of data points, 
especially below 2 eV, it is difficult to discriminate 
between nonresonance and resonance Raman vertex contributions to the 
scattering intensity. However, 
the substantially different resonance behavior of the $zz$ and $xx$ continua
is in qualitative agreement with what one might expect from 
the strong anisotropy of the frequency-dependent Raman 
vertex\cite{Abrikosov74}, which can be 
interpreted as an inverse frequency-dependent effective mass: If one were 
to extrapolate the resonance data to zero frequency,
the value of the out-of-plane continuum scattering efficiency 
would be close to zero and significantly smaller than that of the in-plane
symmetry. This is consistent with a great anisotropy in the  
carrier effective mass, and supported by 
the transport and band structure studies -- that is
$m^{*}_{ab}(\omega = 0)$ $\ll$ $m^{*}_{c}(\omega = 0)$.

From Fig.~3, the pronounced
resonance effect on the $zz$ continuum follows approximately the trend of
the imaginary part of the $c$-axis dielectric 
constant\cite{Tajima93,HLLiu98},
indicating that light couples to 
the $c$-axis continuum {\it via} some intermediate 
excitation states with energy $>$ 3 eV. We believe that
the scattering process in $zz$ geometry is mediated by 
transitions occurring between the CuO$_2$ and Bi-O layers. 
This view is further supported by 
the band theories of Bi-2212 showing that two Bi-O(3) 
bands lie mostly 2-3 eV above the Fermi energy $E_F$ along 
the $\Gamma - Z$ direction\cite{Pickett89}. 
The earlier $c$-axis Raman scattering studies of YBCO indicated that
the resonance behavior results from the modulation of an optical transition 
between CuO$_2$ plane band and chain band\cite{Cooper93}. 
More recently, a theoretical study of the $c$-axis normal-state
Raman scattering also suggested that in some cases
the contribution of interband transitions dominate the 
Raman response\cite{Wu97}.

We now discuss the $c$-axis electronic Raman scattering response
in the superconducting state. As shown in Figs.~1 and~2,
a broad peak in the $c$-axis continuum developed 
below $T_c$ is similar to the data (so-called coherent ``2$\Delta$ peak'') 
observed in the $ab$-plane. 
However, there are some differences between the spectra as well. 
The peak position in $zz$ occurs near 400 cm$^{-1}$ (50 meV). This feature, which
has $A_{1g}$ symmetry, is found at higher frequency than 
that seen in the planar $B_{1g}$ (360 cm$^{-1}$), $B_{2g}$ (345 cm$^{-1}$), and 
$A_{1g}$ (300 cm$^{-1}$) symmetries, but it has a smaller intensity than its 
in-plane counterparts. 
It is worth mentioning that the $B_{1g}$ peak energy
gives 2$\Delta$/k$_B$$T_c$ a value of 6.1, similar to that reported in 
prior work on overdoped Bi-2212\cite{Kendziora95}.
We are presently unaware of any theoretical work that 
would describe the $c$-axis Raman response at $T < {T_c}$. 
It is unclear why the energy scale of the 2$\Delta$ peak associated with the
out-of-plane and the $A_{1g}$ part of the in-plane response
is different. We can say with some certainty that the Raman scattering 
process in both cases creates two quasiparticles. We do not know whether or 
not they are on the same CuO$_2$ plane. Perhaps the difference between the 
two spectra of $A_{1g}$ symmetry occurs because they contain different 
admixtures of same-plane and different-plane pairs of quasiparticles. The two
kinds of pairs could be expected to undergo distinct final state interactions
and screening corrections. The $c$-axis Raman measurements on single-layer
Bi$_2$Sr$_2$CuO$_6$ would augment this study nicely.

Finally, it is of interest to compare our Raman data with the 
recent measurements of the $c$-axis microwave conductivity 
$\sigma_{1c}(T)$\cite{Kitano98}.
In optimally doped Bi-2212, $\sigma_{1c}(T)$ falls rapidly  
below $T_c$, with no sign of the broad maximum observed in the 
$ab$-plane\cite{Shibauchi92}. The behavior of $\sigma_{1c}(T)$ shows 
that the $c$-axis transport 
remains incoherent down to the superconducting transition, and below $T_c$
is better approached as a case of the Josephson tunneling\cite{Kleiner92}. 
The existence of well-defined ``2$\Delta$ peak'' in
our Raman results suggests that three-dimensional superconducting 
coherence is present in overdoped Bi-2212 at low temperatures below $T_c$.

In summary, $c$-axis electronic Raman scattering spectra have been 
investigated for Bi-2212 single crystals. 
Our most significant result is the observation of the $c$-axis
electronic continuum up to very high energies, even though 
the transport in the $c$-direction is incoherent.
The continuum intensity 
dramatically increases using UV photon excitation, suggesting 
that the scattering process in $zz$ geometry is 
dominated by a resonance Raman vertex.
Below $T_c$, there is a
clear low-energy redistribution of the continuum. The appearance of
a 2$\Delta$ peak is a signature of three-dimensional superconducting
coherence. Moreover, the greater value of the peak energy, relative to the 
in-plane cases, signifies that different quasiparticle dynamics are involved. 

\bigskip

We thank S.L. Cooper, K.E. Gray, and A.J. Leggett for helpful discussions.
This work was supported by NSF Grant No. DMR-9705131 (H.L.L.), and 
DMR-9120000 (G.B., M.V.K., and P.G.) 
through the STCS, and DOE-BES W-31-109-ENG-38 (D.G.H.).

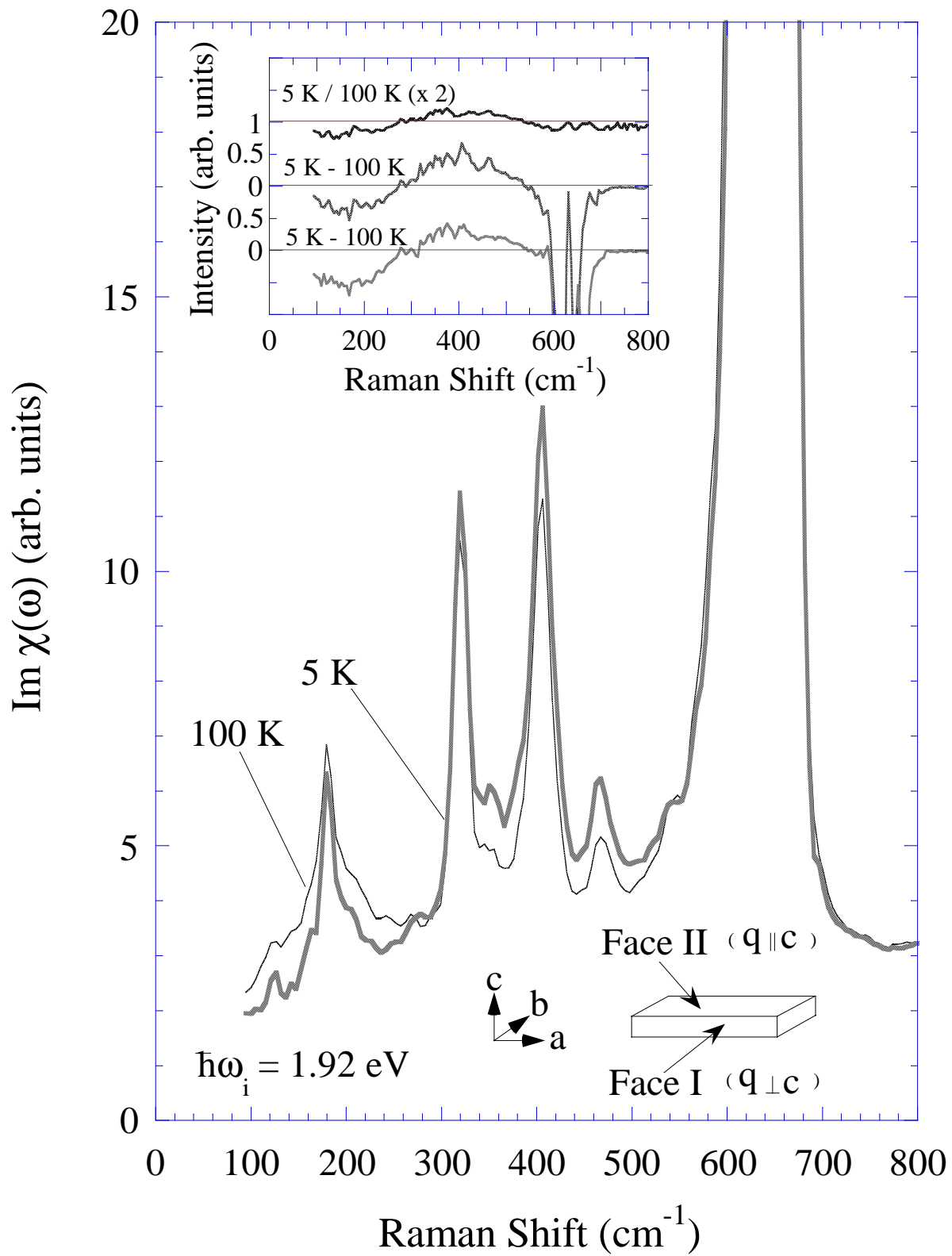
\begin{figure}
\caption{Low-frequency Raman response functions in the $zz$ 
polarization using red (1.92 eV) excitation. Thick line 
denotes the spectra taken at 5 K, and thin line at 100 K.
The upper inset displays the difference between 5 K and 100 K spectra, using 
the procedure described in the text. 
The lower inset shows the sample faces that were used to 
measure Raman spectra.} 
\label{Fig.1}
\end{figure}

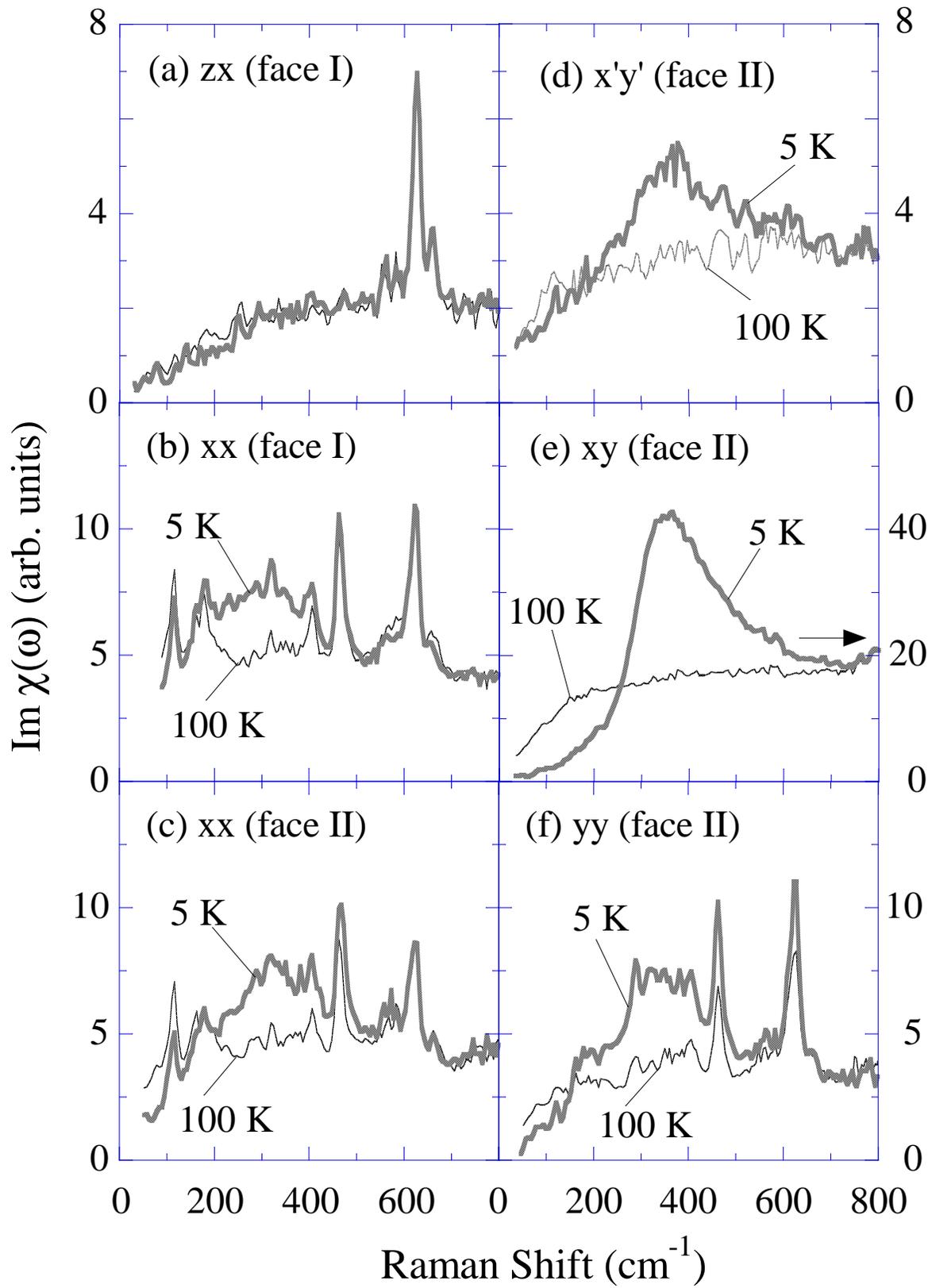
\begin{figure}
\caption{Raman spectra taken with six different polarizations at 
5 K (thick line) and 100 K (thin line).}
\label{Fig.2}
\end{figure}
 
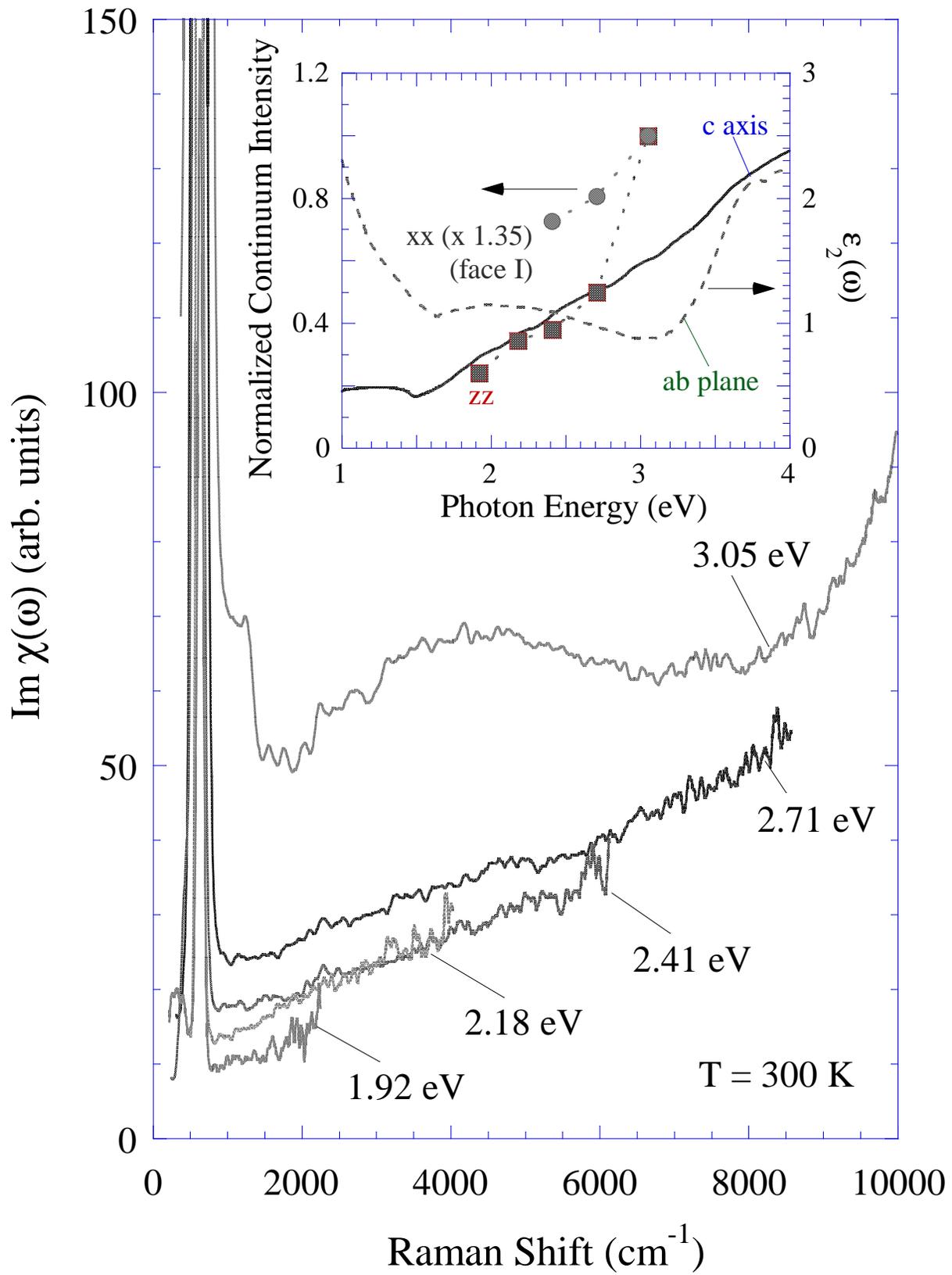
\begin{figure}
\caption{Room-temperature $c$-polarized Raman scattering as a function
of incident photon energy. The inset illustrates the 
normalized resonance Raman excitation profile of the $zz$
(filled square) and $xx$ (filled circle) continua 
compared with the imaginary part of the 
$c$-axis (solid line) and $ab$-plane
dielectric constant (dotted line) [3,25].}
\label{Fig.3}
\end{figure}

\end{document}